# Application of Machine Learning on sequential deconvolution and convolution techniques for analysis of Nuclear Track Detector (NTD) images

Joydeep Chatterjee[a], Rupamoy Bhattacharyya[b], Atanu Maulik[c,1], Kanik Palodhi[a,*]

[a]*Department of Applied Optics and Photonics, University of Calcutta, Kolkata 700106, India*
[b]*Centre for Astroparticle Physics and Space Science, Bose Institute, Kolkata 700 091, India*
[c]*Physics Department, University of Alberta, Edmonton, Alberta, Canada*

**Abstract**

A novel image analysis algorithm as applied to images of Nuclear Track Detectors (NTD) is presented. This process, involving sequential application of deconvolution and convolution techniques, followed by the application of Artificial Neural Network (ANN), is identifying the etch-pit openings in NTD images with a higher degree of success compared to other conventional image analysis techniques.

**Introduction**

Nuclear Track Detectors (NTDs), with their intrinsic high detection thresholds, have often been the detectors of choice when it comes to the detection of highly ionizing particles. This is especially so, when it comes to the search for rare, hypothesized, highly ionizing particles (e.g. Monopoles, Strangelets) against a large background of lighter, low-Z particles [1][2][3].

NTDs like CR-39, Makrofol, PET etc. are basically polymer films. Charged particle passing through them lose energy and if that energy loss is above a certain threshold (different for each material), then the damage to the polymer bonds is sufficient enough for etch-pits to form when such exposed films are subsequently etched in certain chemical solutions. Careful study of the geometry of such etch-pits under optical microscopes reveals vital information about the nature of the particles forming those tracks.

The relative low cost of NTDs, coupled with the fact that they do not require power for their operation, makes them an ideal choice for setting up of large area arrays at remote high altitude locations [1][2].

But one problem that has held back the more widespread adoption of NTDs is that of scanning speed. Automated image analysis has always been the weak point of softwares, with traditional methods using grey level discrimination, eccentricity cuts etc. often coming up short in detecting etch-pits in NTDs, especially in the presence of backgrounds and other structural defects in the detector films. Therefore, scanning NTD films require the painstaking effort of specialists with extensive training. This makes the process prone to human errors and in case of large area arrays, often prohibitively time consuming.

[*] Corresponding author: E-mail address: kpaop@caluniv.ac.in (Kanik Palodhi)

[1]Present address: Istituto Nazionale di Fisica Nucleare, Sezione di Bologna, Bologna 40127, Italy

Recent advances in Machine Learning have promised much better results when it comes to automated image analysis. In a previous work [4], we had employed sequential deconvolution and convolution techniques to obtain significantly improved results compared to current commercially available software when it comes to track identification in NTDs. Subsequently we further improved the method by the application of Artificial Neural Networks (ANN) to automate the setting of thresholds which in the previous case had to be done manually for each image frame. This promises to further speed up the image analysis process. The results of those studies are presented in this paper.

**Proposed method**

Convolution is a common technique quite suitable for comparison of shapes through common area examination. Mathematically, convolution of two 2-dimensional (2D) functions, (here, shapes) $a(x, y)$ and $b(x, y)$ is given below [5].

$$f_1(x, y) = a(x, y) * b(x, y) = \iint_{-\infty}^{\infty} a(x - x_0, y - y_0) b(x_0, y_0) \, dx_0 dy_0 \tag{1}$$

In Fourier domain:

$$F_1(u, v) = A(u, v) B(u, v) \tag{2}$$

Considering the functions $A(u, v)$, $B(u, v)$ and $F_1(u, v)$ to be the Fourier transforms of the functions $a(x, y), b(x, y)$ and $f_1(x, y)$, respectively, (where $x$ and $y$ be the coordinates in the image space and $u$ and $v$ be the corresponding spatial frequency coordinates in the Fourier space).

Depending upon the closeness of the shapes i.e. $a(x, y)$ and $b(x, y)$, a 2D resultant function with a sharp peak is obtained after convolution. This is shown in the figure 1 below for a circle and ellipse with the circle used as the convolution mask, the shape convolved with each of the shapes.

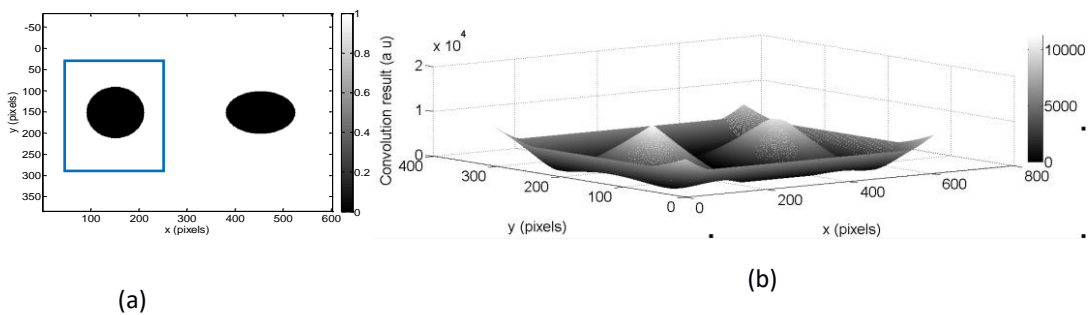

(a)

(b)

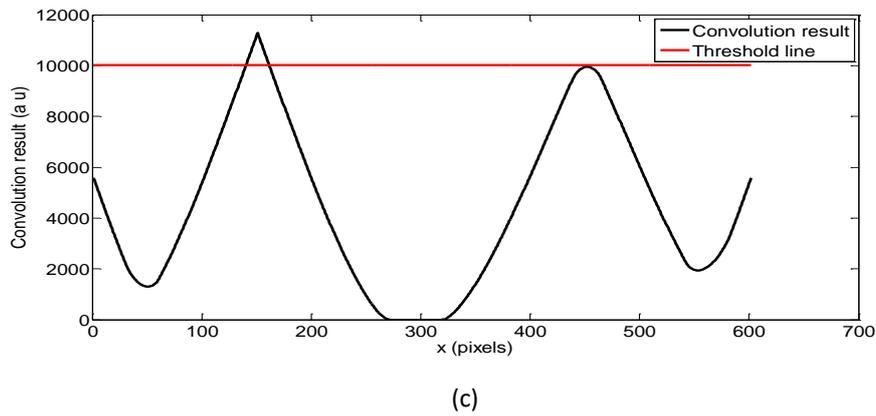

(c)

Figure 1: (a) Two shapes circle and ellipse with the circle in blue box also serving as the convolution mask; (b) Result of convolution with the circle being used as the circular convolution mask; (c) Plot along the middle row along the length of the Fig. 1(b) with the red line as the threshold line as the result of convolution where the difference in peak heights is 1339 a. u.

Deconvolution is essentially, an "inverse" operation to convolution. From equation (2), the deconvolution of $f_1(x,y)$ and $b(x,y)$ is given by:

$$A(u,v) = \frac{F_1(u,v)}{B(u,v)} \qquad (3)$$

In the Fig. 1(a), the circle and the ellipse are the objects. The circle, shown with the blue rectangle is also the convolution mask used. Fig. 1(b) shows the 2D mesh profile after the convolution. Fig. 1(c) is the plot of the mid array along the length of Fig. 1(b) where the difference between the peak heights for the peaks obtained for circle and the ellipse are distinct. The peak height difference is 1339 a. u. for this case and by setting a proper threshold line shown in Fig. 1(c) as a red line, the peak for the circle can be distinguished from that obtained for the ellipse.

In our previous work, Palodhi *et al* [4] has shown that in image analysis of NTDs deconvolution followed by convolution results in higher peaks compared to only convolution if applied to same shapes as shown in figure 2, where a circular mask is used for two of these operations on itself, first only convolution and in the second case, sequential deconvolution and convolution. Clearly, two 1D curves through both the peaks represent the enhanced sensitivity of the second operation.

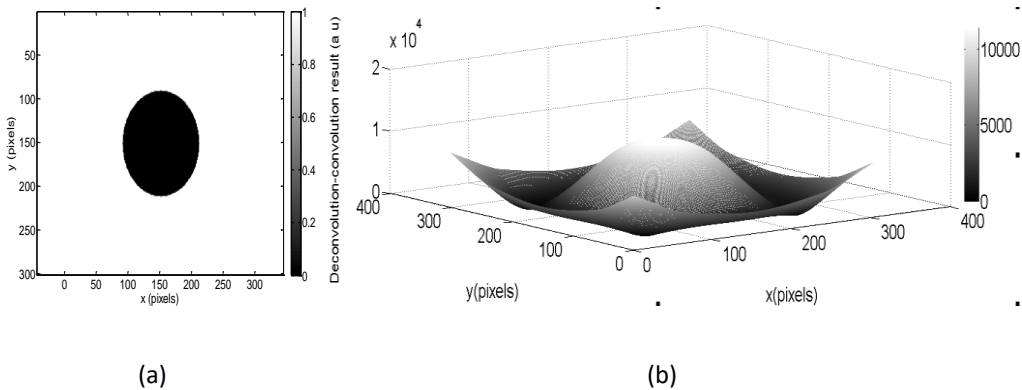

(a)　　　　　　　　　　　　　　　　(b)

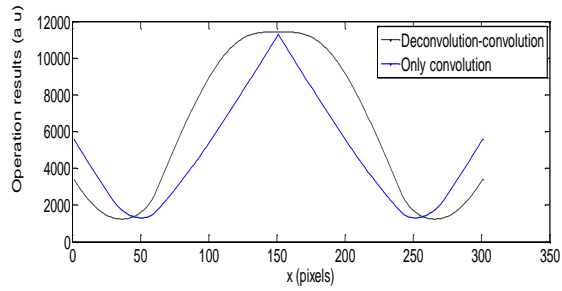

(c)

Figure 2. (a) Object, (b) Result for deconvolution followed by convolution and (c) Graph for deconvolution followed by convolution and only convolution with the peak height difference 120 a. u.

In the context of NTD images, this means that if a suitable mask is designed from the existing etch-pit openings in an image, after sequential deconvolution-convolution, there will be several peaks generated due to presence of etch-pits and other artefacts such as scratches, surface defects etc. Among them, the sharp peaks will be due to etch-pits. The aim is to isolate these sharp peaks from the background with proper thresholding.

Previously, Palodhi *et al* [4] used manual thresholding for this purpose. Although it worked well for the images tested, it was image specific and each time a new threshold had to be determined. The main goal of this present paper is to automate this thresholding process using Artificial Neural Networks (ANN) [6]. Unlike the previous case [4], this method proposed here uses a Gaussian mask for deconvolution with the NTD image and then uses a circular mask for convolution with the result. This will result in some sharp peaks at the track positions along with some undulations in the background.

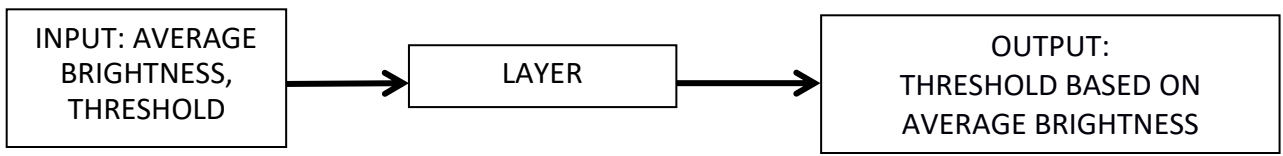

Fig. 3: Neural network model used for NTD

Neural networks are inspired by the nervous system of the living beings. Similar to the nervous system, information is carried through neurons to the processing units. The structure of the neural network used in this paper is given below in Fig. 3. The neural network is of feed-forward architecture where the dataflow is in the forward direction. [6]

The details of the neural network have been given below in Table 1:

Table - 1

| Input | Parameters: Average brightness after deconvolution followed by convolution, threshold value |
|---|---|
| Hidden layer | One hidden layer used |
| Output | Threshold based on the average brightness after deconvolution and convolution. |

Mathematically,

$$y_i = f\left(\sum_{j=0}^{N} w_{ij} x_j\right) \quad (4)$$

Here, $w_{ij}$, $x_j$ and $y_i$ are the weight, input, and output for every neuron respectively [7].

The average intensity of the raw figure obtained after deconvolution and convolution operations and the manual threshold values for the training images are used as the training parameters of the network. The trained neural network is used for predicting the threshold taking the average intensity of the deconvolution and convolution results for the test image and the isolated peaks are counted using morphological techniques.

**Experiment**

A set of films of CR-39 and PET, of thickness 700 micron and 100 micron respectively, were exposed to 3.9 MeV/nucleon $^{32}$S beams from the pelletron accelerator at IUAC, New Delhi. Another set of NTD foils were given open air exposure at a high altitude (~2200 m a.m.s.l) location at Darjeeling, in the Eastern Himalayas. The exposed films were then etched in 6.25 N NaOH solutions at 55 °C for PET and 70 °C for CR-39 films for 3 hrs. The etched films were then studied under a Leica DMR optical microscope and images of the NTD surface captured for subsequent analysis.

The obtained images were first subjected to a image processing algorithm as described previously [4]. First, based on the biggest track size, a Gaussian mask and a circular mask were generated. Then, the obtained images were divided into training and test image sets. Following common conventions, typically 75 per cent images of the entire dataset were taken as the training dataset and the rest as test images as shown in Table 2. . Each of the training images were first deconvolved with the above mentioned Gaussian mask. The resulting images were then convolved with the circular mask generated earlier. The resulting images contain distinct high-peaks at the positions of the tracks. Now for each training image, proper manual threshold limit was set in such a manner such that only the peaks which are higher than the threshold values are present in the thresholded image whereas the background noise, artefacts etc. are suppressed. The average intensity of the raw figure obtained after deconvolution and convolution operations are also determined for each of the training images.

The average intensity values and the proper manual threshold values set for the corresponding images are the training parameters for the neural network to be used. After being trained with these input values, the neural network will be ready to make prediction of the threshold level to be set. Hence, one test image is first deconvolved with the previously used Gaussian mask and then with the previously applied circular mask and the average intensity value of the resulting image is determined and fed to the neural network to get the predicted threshold values. After setting the suggested threshold value the isolated peaks are identified, marked and counted with simple morphological operations and the counting numbers are shown on the tracks. Three separate neural networks were used. One for 0° angle of incidence of the ions from the accelerator, another for 30° incidence and the third for field (Darjeeling) images and each of them produced successful results as shown below.

**Results and discussions**

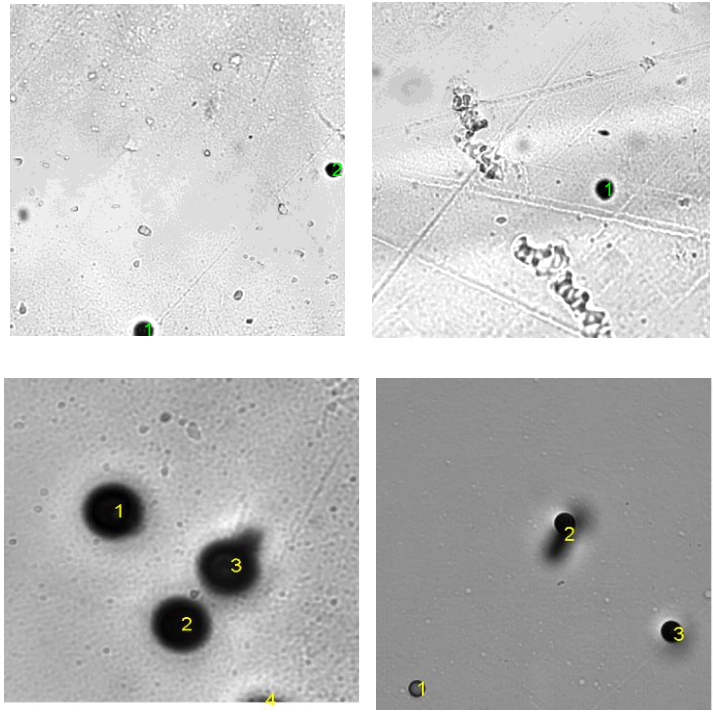

Fig. 4: Counting results for 0° incidence in the accelerator

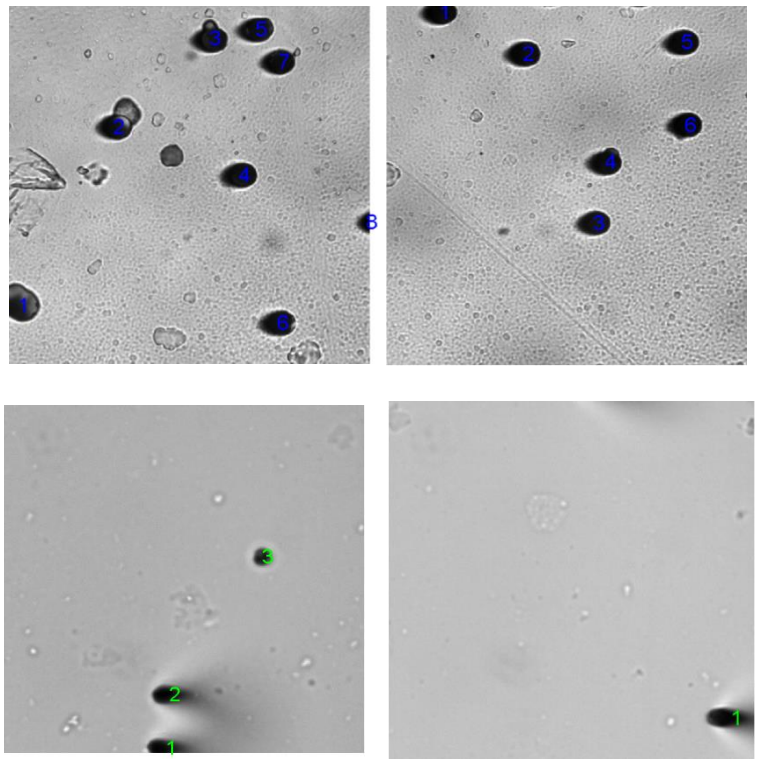

Fig. 5: Counting results for 30° incidence in the accelerator

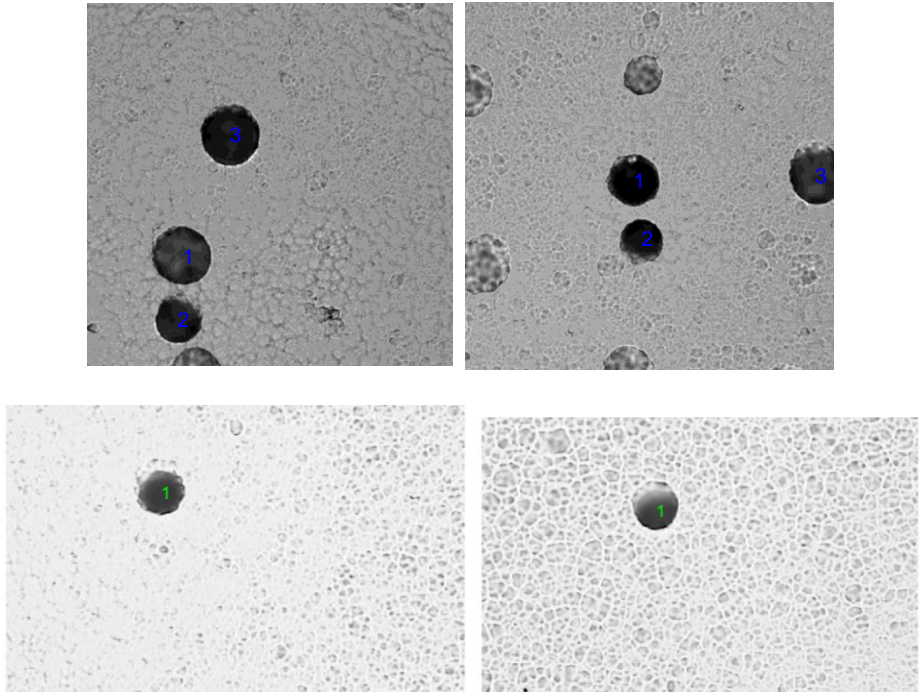

Fig. 6: Counting results for field (Darjeeling) data

From the above figure, it can be seen that the tracks have been counted very efficiently. The success rates are about 94%, 93% and 90% for the three types of images mentioned above. Even some partially visible tracks are also identified and counted. The effects of scratches, artefacts etc. have been dealt with quite efficiently. The algorithm is fast, simple and not much computationally intensive.

Table - 2

| Image type | Total no. of images | Training images | Test images | Accuracy |
|---|---|---|---|---|
| Accelerator $0^o$ | 60 | 45 | 15 | 94% |
| Accelerator $30^o$ | 60 | 45 | 15 | 93% |
| Field (Darjeeling) | 50 | 37 | 13 | 90% |

We have compared the efficacy of this Neural Network based approach with other techniques. For example, if the estimate of the threshold is made employing a ratio based linear estimation approach, using average brightness values and manual threshold values from the training images, then, as shown in Fig. 7 (a), the track count (11) can be erroneous as some of the background defects are included in the track count. In contrast, the Neural Network based approach applied to the same image frame as shown in Fig. 7(b), has correctly identified all 6 tracks, even including the ones which are partially seen, while ignoring all background defects.

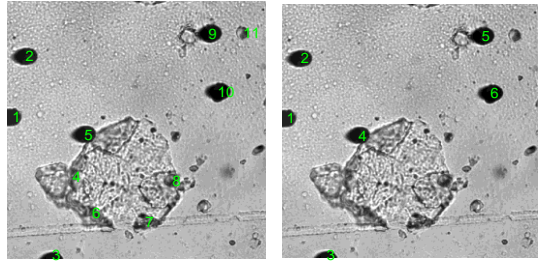

Figure 7: (a) Ratio based linear estimation approach counts 11 tracks, (b) The neural network based approach correctly counts 6 tracks, ignoring the defects

This shows that the Neural Network based method produces more accurate results compared to other approaches.

**Conclusion**

The image analysis process described above, combining deconvolution and convolution techniques and Artificial Neural Network (ANN) is producing excellent results when it comes to NTD track image analysis. This method promises a much faster and efficient automated image analysis system for NTD images, thereby removing a serious technological bottleneck and facilitating a more widespread deployment of NTDs.

**Acknowledgement**

One of the authors, AM, would like to thank Dr. Laura Patrizii of INFN, Bologna, Italy and Dr. James Pinfold of University of Alberta, Edmonton, Canada for financial support and encouragement. Authors would also like to thank the staff at IUAC, New Delhi, India for their help during the exposure of the NTDs. They would also like to thank the staff at Bose Institute, Kolkata, India for their help during the etching and scanning process. JC and KP would like to thank University of Calcutta for providing the lab support.